\documentclass[aps,prl,showpacs,superscriptaddress,twocolumn]{revtex4-1}
\usepackage[dvips]{graphicx}
\usepackage{natbib}
\usepackage{hyperref}
\usepackage{amsmath,amsfonts,amssymb}
\usepackage{multirow} 
\usepackage{times}

\newcommand{\cc}[1]{{#1}^{*}} 

\newcommand{\rme}{ \mathrm{e} }
\newcommand{\rmi}{ \mathrm{i} }
\newcommand{\rmd}{ \mathrm{d} }
\newcommand{\Beta}{ \mathrm{B} }

\newcommand{\artanh}{\mathop{\mathrm{artanh}}}
\newcommand{\tr}{\mathop{\mathrm{Tr}}}
\newcommand{\sgn}{\mathop{\mathrm{sgn}}}
\newcommand{\erfc}{\mathop{\mathrm{erfc}}}
\newcommand{\im}{\mathop{\mathrm{Im}}}

\begin{document}

\title{Truncations of Random Orthogonal Matrices}

\author{Boris A. Khoruzhenko}
 \affiliation{Queen Mary University of London, School of Mathematical Sciences,
 London E1 4NS, UK}
\author{Hans-J{\"u}rgen Sommers}
 \affiliation{Fakult\"at f\"ur Physik, Universit\"at Duisburg-Essen, 47048 Duisburg, Germany}
\author{Karol {\.Z}yczkowski}
 \affiliation{Smoluchowski Institute of Physics,
Jagiellonian University, ul. Reymonta 4, 30-059 Krak{\'o}w, Poland}
\affiliation{Centrum Fizyki Teoretycznej, Polska Akademia Nauk,  Al. Lotnik{\'o}w 32/44, 02-668 Warszawa, Poland}
\date{12 August, 2010}

\begin{abstract}
Statistical properties of non--symmetric real random matrices of size $M$, obtained as truncations of random orthogonal $N\times N$ matrices are investigated. We derive an exact formula for the density of eigenvalues which consists of two components: finite fraction of eigenvalues are real, while the remaining part of the spectrum is located inside the unit disk symmetrically with respect to the real axis. In the case of strong non--orthogonality, $M/N=$const, the behavior typical to real Ginibre ensemble is found. In the case $M=N-L$ with fixed $L$, a universal distribution of resonance widths is recovered.
\end{abstract}

\pacs{05.40.-a, 02.50.-r, 75.10.Nr}

\maketitle

Random unitary matrices \cite{Dyson1962}
are used in numerous physical applications
including chaotic scattering, conductance in mesoscopic
systems \cite{Beenakker1997} or statistical properties of periodically driven
quantum systems  \cite{Ha06}.
In several applications one needs to restrict this class to real,
orthogonal matrices, which are assumed to be distributed uniformly
with respect to the Haar measure on the orthogonal group.
This is the case while describing quasiparticle excitations in metals and superconductors
\cite{AZ97,DBB10},
or quantum maps performed on real quantum states \cite{BZ06}. In applications one is often led to consider square truncations of random unitary matrices of large matrix dimension $N$. These matrices have been used to describe quantum systems with absorbing boundaries  \cite{CMS99} and found applications in various physical problems including optical and semiconductor superlattices \cite{GKK02}, problems of quantum conductance \cite{Fo06}, distribution of resonances for open quantum maps \cite{FS03,SJ05,NZ07,PCW09}. The operation of truncation does not preserve unitarity and as a consequence the eigenvalues move inside the unit disk in the complex plane, see, e.g., \cite{Ha06,FK09}, and the emerging eigenvalue statistics appear to be universal \cite{FS03}.

The aim of this work is to present a comprehensive study of truncations of {\sl random orthogonal matrices}. We consider an ensemble of $N\times N$ real orthogonal matrices $Q$ with flat matrix distribution on the orthogonal group. The goal of this paper is to compute analytically the full distribution of eigenvalues of the top $M\times M$ sub-matrix $A$ of $Q = {A \ B \choose C \ D}$ for any values of $M$ and $N$\cite{ftn}. This then gives the eigenvalue correlation functions in a closed form and we investigate two asymptotic regimes of direct physical importance. For simplicity we consider $M$ even, however, our results can be generalized to odd dimensions as well.


The orthogonality condition for the first $M$ columns of $Q$, $A^T\!A+C^T\!C=1$, defines a manifold in the space of real $N\times M$ matrices which can be identified with the flag manifold $O(N)/O(L)$ where $L=N-M$. Correspondingly, the density of distribution of $A,C$ is
\vspace*{-1ex}
\[
P(A,C)=\frac{\text{vol}\, O(L)}{\text{vol}\, O(N)}\ \delta(A^T\!A+C^T\!C-1),
\]
where $\text{vol}\, O(N)=2^N\, \prod_{j=1}^N {\pi^{j/2}}/{\Gamma (j/2)}$ is the volume of the orthogonal group $O(N)$ \cite{BZ06}.
The density function of $A$ is $P(A,C)$ integrated over $C$. For $L\ge M$ the $C$-integral can be performed by scaling $C\to C\sqrt{1-A^T\!A}$. The corresponding Jacobian is $|\det(1-A^T\!A)|^{L/2}$ and from the $\delta$-function one gets another factor $|\det(1-A^T\!A)|^{-(M+1)/2}$. The remaining integral  yields the volume of $O(L)/O(L-M)$, leading to\cite{Fo06}
\begin{equation}\label{eq:b3}
P(A)=\frac{\text{vol}\, O(L)\, \text{vol}\, O(L)}{\text{vol}\, O(N)\, \text{vol}\, O(L\!-\!M)}\, \det(1-A^T\!A)^{\frac{L-M-1}{2}}
\end{equation}
If $L\!<\!M$ the probability density function of $A$ contains singular ($\delta$-function) terms due to finite mass of the boundary of the matrix ball $A^T\! A \le 1$ which supports the distribution of $A$.

By setting $M=1$ in Eq.~(\ref{eq:b3}) one recovers the distribution $P(A)\propto (1-A^2)^{\frac{N-2}{2}}$ of a single entry of $Q$. Obviously, for $N$ large $P(A)$ can be approximated by a Gaussian. This is also true for matrix blocks if the size of truncation $M\ll N$. Eigenvalue correlations in such a Gaussian regime are now accessible following the recent progress \cite{Kanzieper2005,FN07,Som07,BS09,FM10} made for the real Ginibre ensemble \cite{Gi65}.
In this paper we investigate two non-Gaussian regimes: (i) strong non-orthogonality $M,L \propto N$; and (ii) weak non-orthogonality $M\propto N$ and $L\ll N$. In the former we again recover Ginibre eigenvalue correlations while the latter yields eigenvalue statistics from a different universality class \cite{FS03}.

We shall obtain the joint distribution of eigenvalues of $A$ directly from $P(A,C)$ reducing the general case to $M=2$. Thus, it is instructive to consider first $M=2$. In this case, one can bring $A$ by a rotation to the form $A= {\phantom{-}\lambda_1 \, \delta \choose -\delta \, \lambda_2}$ with real $\lambda_{1,2}$ and $\delta$ from which the eigenvalues $z_i$ of $A$ can be recovered via the relations $\det A = \lambda_1 \lambda_2+\delta^2$ and $\tr A = \lambda_1+\lambda_2$ \cite{Som07}. The matrix $A$ can have either two real eigenvalues in which case we choose $z_1>z_2$ or two complex conjugate eigenvalues in which case we choose $\im z_1 > \im z_2$. The joint distribution $\rmd \mu (z_1,z_2)$  for $L\ge 2$ in each of these two cases can be obtained from Eq.~(\ref{eq:b3}) by making use of $\det (1- A^T\!A)=(1-z_1^2)(1-z_2^2)-4\delta^2$ and integrating out $\delta$ subject to the constraints $4\delta^2 \le (1-z_1^2)(1-z_2^2)$ and $4\delta^2\ge -(z_1-z_2)^2$ arising due to the chosen parameterization of $A$, see \cite{Som07}, and the positivity of $1-A^T\! A$. The resulting expression can be presented in a form that encodes both cases (real and complex):
\vspace*{-2ex}
\begin{equation}\label{eq:b4}
\rmd \mu (z_1,z_2) = (z_1-z_2)\, f(z_1)f(z_2)\, dz_1 \wedge dz_2
\end{equation}
with ($z=x+\rmi y$)
\begin{equation}\label{eq:b5}
f^2(z)=\frac{L(L-1)}{2\pi}\ |1-z^2|^{L-2}\!\! \int_{\frac{2|y|}{|1-z^2|}}^1\!\!
\rmd u\,  (1-u^2)^{\frac{L-3}{2}} .
\end{equation}
If the eigenvalues of $A$ are real then the integral in
(\ref{eq:b5}) yields a Beta function. By integrating $\rmd \mu
(z_1,z_2)$ over the triangle $-1\!<\!z_2\!<\!z_1\!<\!1$ one gets the probability
$p^R_{2,L}$ for $A$ to have two real eigenvalues:
$p^R_{2,L}=\Beta(L/2+1/2,L)/\Beta(L/2,L\!+\!1/2)$. In the limit
$L\to\infty$ this converges to $1/\sqrt{2}$ which, as expected, is the value for
the real Ginibre ensemble.

One recovers $\rmd \mu (z_1,z_2)$ for $L=1$ by letting $L\to 1$ in (\ref{eq:b5}). In this limit $f^2(z)=(2\pi |1-z^2|)^{-1}$. As a simple check one can evaluate the probability $p^C_{2,L=1}$ for $A$ to have two complex eigenvalues. In this case $z_1=\cc{z_2}$ so that $(z_1-z_2)dz_1 \wedge dz_2=4y\rmd x \rmd y$ in (\ref{eq:b4}). Integrating $\rmd \mu (z_1,z_2)$ over the upper half of the unit disk $|z|<1$ one gets $p^C_{2,1}=1-2/\pi=1-p^C_{2,1}$, in agreement with the above result for $p^R_{2,L}$.

We now proceed with the general case of $M$ even. By an orthogonal transformation ${\cal O}$ drawn from $O(M)/O(2)^{M/2}$ one can bring $A$ to upper triangular form $A={\cal O} (\Lambda +\Delta) {\cal O}^T$, where $\Lambda$ is block diagonal with $2\times 2$ blocks $\Lambda_I$ in the diagonal and $\Delta$ is block triangular with $2\times 2$ blocks $\Delta_{IJ}$ ($I<J$) above the diagonal. Integrating $P(A,C)$ over ${\cal O}$, $\Delta$ and $C$ we obtain the distribution of $\Lambda$,
\begin{eqnarray}\label{eq:b6}
P(\Lambda)&=&(4\pi)^{-M/2}\, c_{M,L} \, \Delta^{\prime}(\Lambda)
\times \\
\nonumber && \int \rmd  \Delta \! \int \rmd C\,  \delta((\Lambda^T+\Delta^T)(\Lambda+\Delta) +C^T\!C-1)
\end{eqnarray}
with $c_{M,L}\!=\!\text{vol} O(L)\text{vol}O(M)/\text{vol}O(L\!+\!M)$.
Here $\Delta^{\prime}(\Lambda) = \Delta (\Lambda)/\prod_{I} \Delta (\Lambda_I)$ is the Jacobian of the transformation from $A$ to ${\cal O}$, $\Delta$ and $\Lambda$, see \cite{KS09}, and $\Delta (\Lambda)=\prod_{i<j}(z_i-z_j)$ is the Vandermonde determinant of the eigenvalues $z_j$ of $\Lambda$. Thus $\Delta^{\prime}$ does not count the pairs $i<j$ from the diagonal blocks.

Writing $C$ as $C=[C_1, \ldots, C_{M/2}]$ in terms of $L\times 2$
sub-blocks $C_I$, we have for $I<J$ the equations
\begin{equation}\label{eq:b7}
\Lambda_I^T\! \Delta_{IJ} +\sum_{K<I}\Delta_{KI}\Delta_{KJ} = - C_I^T\! C_J
\end{equation}
due to the $\delta$-function in (\ref{eq:b6}), which start with $\Lambda_1^T\!\Delta_{1J}=-C_1^T\!C_J$ for $J\ge 2$, $\Lambda_2^T\!\Delta_{2J} + \Delta_{12}^T\! \Delta_{2J} =-C_2^T\!C_J$ for $J\ge 3$, etc. Because of the triangular structure of $\Delta$ these equations can iteratively be solved for $\Delta_{IJ}$. Thus integration over $\Delta_{IJ}$ can be performed with a Jacobian $\prod_{I<J} |\det \Lambda_I|^{-2}$. There remains integration over $C_J$:
\[
\int \prod_{J=1}^{M/2} [\rmd C_J\, \delta (\Lambda_J^T\! \Lambda_J + \sum_{K<I}\Delta_{KI}\Delta_{KJ} + C_J^T\! C_J-1 )].
\]
From Eqs.~(\ref{eq:b7}) one sees that $\Delta_{IJ}$ for $I<J$ is linear in $C_J$ and depends otherwise on $C_K$ for $K<J$. Thus we have an integral
$
\int \prod_{J=1}^{M/2} \rmd C_J\, \delta (\Lambda_J^T\! \Lambda_J + C_J^T\!X_J C_J-1 )
$
where $X_J$ is a linear operator acting on $C_J$ per matrix multiplication. The scaling $C_J\to (1/\sqrt{X_J}) C_J$ results in a Jacobian that cancels the Jacobian $\prod_{I<J} |\det \Lambda_I|^{-2}$ from the previous integration over $\Delta_{IJ}$ and we end up with
\[
\frac{P(\Lambda)}{c_{M,L}}=  
\frac{\Delta^{\prime}(\Lambda)}{(4\pi)^{M/2}}
\int \prod_{J=1}^{M/2} \rmd C_J\, \delta (\Lambda_J^T\! \Lambda_J + C_J^T\! C_J-1 ).
\]
where $\Lambda_J$'s are $2\times 2$ and $C_J$'s are $L\times 2$. Now we proceed calculating $P(\Lambda)$ using our result for $M=2$. Performing integration over $C_J$ yields
\[
\frac{P(\Lambda)}{c_{M,L}}=\frac{\Delta^{\prime}(\Lambda)}{(4\pi)^{M/2}} \prod_{J=1}^{M/2} \frac{\text{vol}\, O(L)}{\text{vol}\, O(L-2)}\, \det (1-\Lambda_J^T\! \Lambda_J)^{\frac{L-3}{2}},
\]
and we obtain the joint (full) distribution of eigenvalues of $A$:
\begin{equation}
\label{eq:b8}
      {\rm d}\mu(z_1,...,z_M)= {\cal Z}_{M,L}^{-1}
      \! \prod_{ 1\le i < j \le M}\!\!\! (z_i-z_j)\,
      \prod_{j=1}^M f(z_j)\, dz_j
\end{equation}
with ${\cal Z}_{M,L}^{-1}= [L!/(2\pi)^L]^{M/2}\! c_{M,L}$, $dz_j\!=\!dx_j+\rmi dy_j$ and ordering $z_1>z_2> \ldots >z_M$ if all eigenvalues are real, $z_1=\cc{z_2}$, $\im z_1>0$, $z_3>z_4> \ldots >z_M$ if two eigenvalues are complex conjugate and so on.  Eqs.~(\ref{eq:b8}), (\ref{eq:b5}) represent our first main result.

Eq.~(\ref{eq:b8}) is of the same form as the corresponding one in the real Ginibre ensemble \cite{Som07} but with a different weight function $f(z)$. Following \cite{Som07} one can obtain the eigenvalue correlation densities in terms of a skew-symmetric kernel
\[
{\cal K} (z_1, z_2 ) =   \sum_{k=1}^M \sum_{l=1}^M  ({\cal A}^{-1})_{kl} z_1^{k-1} z_2^{l-1}
\]
where ${\cal A}_{kl} =  \int\int\! {\rm d}^2 z_1{\rm d}^2z_2\, {\cal F}(z_1, z_2)\, z_1^{k-1} z_2^{l-1}$ with $\rmd^2 z =\rmd x\rmd y$ and
\begin{eqnarray*}\label{eq:b11}
\lefteqn{
{\cal F} (z_1, z_2 )  = f(z_1) f(z_2)  \times } \\ \nonumber && [2\rmi\, \delta^{(2)}(z_1-\cc{z_2}) \sgn(y_1) + \delta(y_1) \delta(y_2) \sgn(x_2-x_1)],
\end{eqnarray*}
with $\delta^{(2)}(z_1-\cc{z_2})=\delta(x_1-x_2)\delta(y_1+y_2)$. For example, the one-point density is given by an integral
\begin{equation}\label{eq:b10}
R_1 (z) = \int {\rm d}^2z_2 \, {\cal F} (z, z_2) {\cal K} (z_2,z).
\end{equation}
Higher order densities $R_n$ are given by a Pfaffian involving ${\cal F}$  and ${\cal K}$ \cite{Som07}. Expressing the kernel ${\cal K}$ in terms of  skew-orthogonal polynomials with respect to weight function ${\cal F}$ leads to an alternative Pfaffian representation for $R_n$ \cite{Kanzieper2005,FN07}.

Evaluating the kernel ${\cal K}$ (or, equivalently \cite{AKP10}, the corresponding skew-orthogonal polynomials) in a closed form is an important step on the way to obtaining eigenvalue statistics \cite{FN07}. In our case the kernel can be found by exploiting its relation to averages of the characteristic polynomials \cite{Som07,APS09} of truncations of orthogonal matrices of smaller dimension:
\[
{\cal K} (z_1,z_2)= (z_1-z_2)\, \langle \det(z_1-\tilde A)(z_2-\tilde A) \rangle_{\tilde A}
\]
where the matrices $\tilde A$ are square truncations of size $M-2$ of random orthogonal matrices of size $N-2$. Due to the invariance of the distribution of $\tilde A$ the above average is effectively an average over the eigenvalues of ${\tilde A}^T\!{\tilde A}$:
\[
\langle \det(z_1-\tilde A)(z_2-\tilde A) \rangle_{\tilde A} = \sum_{m=0}^{n} (z_1z_2)^{m} \frac{\langle \epsilon_{n-m} ({\tilde A}^T\!{\tilde A}) \rangle_{\tilde A}}{\epsilon_{n-m} (1)},
\]
where $n=M\!-\!2$ is the size of $\tilde A$ and $\epsilon_m (X)$ is the $m$-th elementary symmetric function in eigenvalues of $X$. The integral  $\langle \epsilon_m ({\tilde A}^T\!{\tilde A}) \rangle_{\tilde A}$ can be reduced \cite{FK07} to Selberg's integral \cite{Me91},
yielding the kernel in a closed form,
\begin{equation}  \label{eq:b9}
{\cal K}(z_1,z_2)= (z_1-z_2)\, \sum_{m=0}^{M-2}\frac{(L+m)!}{L!\, m!}\,  (z_1z_2)^m,
\end{equation}
which is our second main result. The truncated binomial series on the rhs can be expressed \cite{KS09} in terms of an incomplete beta function which comes in handy for asymptotic analysis of eigenvalue statistics in the limit of large $M$.

The one-point eigenvalue density (\ref{eq:b10}) is composed of two parts: $R_1 (z) = \rho_C(z) + \delta(y)\rho_R(x)$, where $\rho_C(z) = 2 f(z)^2 |{\cal K} (z,\cc{z})|$ is the density of complex eigenvalues and
\vspace*{-2ex}
\begin{equation}\label{eq:b12}
\rho_R (x) = \int_{-1}^1 \rmd x_2\,  \sgn (x_2-x) {\cal K} (x_2,x)
f(x_2)f(x)
\end{equation}
is the density of real eigenvalues of truncated orthogonal matrices, with the normalization $\int_{|z|<1} R_1(z) \, \rmd^2 z=M$.

We shall first look at real eigenvalues of $A$. Integrating $\rho_R(x)$ over $-1 < x < 1 $ one obtains the expected total number $N_R$ of real eigenvalues,
\[
N_R =
1+\frac{L}{2}\int_0^1 \!\! \frac{\rmd s}{s^{L+1}}\
  I_{s^2}\left({L}/{2}, {1}/{2}\right) \  I_{\frac{2s}{1+s}}\left(L+1,M-1\right),
\]
where $I_x(a,b)=\int_0^x t^{a-1}(1-t)^{b-1} {\rm d} t/ \!B(a,b)$ is an incomplete Beta function. For large matrix dimensions ($M\gg 1$) $N_R$, in the leading order, is described by a simpler expression:
\[
N_R \simeq \frac{2\artanh \sqrt{M/N}}{B(L/2, 1/2)}\, .
\]
Depending on $L$ this leads to different scaling laws for $N_R$. In the limit of strong non-orthogonality when both $M, L\propto N$, the number of real eigenvalues $\propto \sqrt{M}$ which is characteristic of the real Ginibre ensemble \cite{Edelman1994}. On the other hand, in the limit of weak non-orthogonality when $L\ll M$, $N_R$ grows logarithmically with $M$, $N_R \simeq (\log M)/{B(L/2, 1/2)}$.

Eq.~(\ref{eq:b12}) can be transformed to a form,
\begin{align} \nonumber
\rho_R (x) = &
\frac{1}{B({L}/{2},{1}/{2})}\frac{I_{1-x^2}(L+1,M-1)}{1-x^2}
+  \\ \nonumber  &
\frac{(1-x^2)^{\frac{L-2}{2}}|x|^{M-1}}{B({M}/{2},{L}/{2})}\,
I_{x^2}((M-1)/2,(L+2)/2),
\end{align}
revealing the Artanh Law of the distribution of real eigenvalues for
large matrix dimensions $M$  ($\mu = M/N$):
\[
\rho_R (x) \simeq \frac{1}{B(L/2, 1/2)}\, \frac{1}{1-x^2} \quad
-\sqrt{\mu} < x < \sqrt{\mu}.
\]
\begin{figure}
\includegraphics[width=0.6\linewidth]{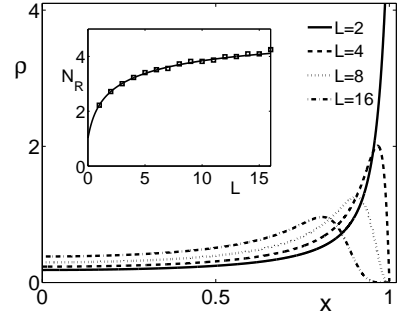} 
\caption{\label{Fig1} Rescaled density of real eigenvalues $\rho=\frac{\rho_R}{N_R}$ for
$M=32$ and $L=2,4,8,16$. The inset shows the number of real eigenvalues $N_R$
for $M=32$ as a function of $L$ obtained from samples of $10^3$ realizations (squares)
and analytical prediction (solid line)}
\end{figure}

In the limit of strong non-orthogonality $\mu <1$ and $\rho_R (x)$
is supported strictly inside the interval $(-1,1)$, vanishing at a
Gaussian rate at the boundaries $x=\pm \sqrt{\mu}$. In the limit of
weak non-orthogonality $\mu \to 1$ and $\rho_R (x)$ becomes singular
at the boundaries $x=\pm 1$ of the real eigenvalue support.
Correspondingly, real eigenvalues accumulate close to
$x=\pm 1$. Setting $x=1 - u/M$ and taking the limit $M\to \infty$
one obtains the density profile near the accumulation point $x=1$:
$\rho_R (x)/M\simeq p(u)$ with
\begin{eqnarray*}
p(u)& =& \frac{u^{\frac{L}{2}-1}\rme^{-u}}{2\Gamma(L/2)} (1-u^{\frac{L}{2}+1}\gamma^{*}({L}/{2}+1, u) ) + \\ \nonumber  & &
\frac{(2u)^L}{B(L/2, 1/2)}\gamma^{*}(L+1, 2u) \, ,
\end{eqnarray*}
where $\gamma^*(n,x)=({x^n \Gamma (n)})^{-1}\! \int_0^x
t^{n-1}\rme^{-t}\rmd t$ is an incomplete Gamma function.
For small $u$,
$p(u) \simeq (2\Gamma(\frac{L}{2}))^{-1} u^{\frac{L}{2}-1}$,
showing a
transition, as $L$ increases, in nature of the boundary points $x=\pm 1$ from
`attractive' ($L=1$) to `repulsive' ($L\ge 4$), see Fig.~\ref{Fig1}. On the other hand, for
large $u$,
$p(u)\simeq (B({L}/{2},{1}/{2})u)^{-1} $,
exhibiting a
heavy tail which manifest itself in the $\log M$ asymptotics for
$N_R$.

\begin{figure}
\includegraphics[width=0.8\linewidth]{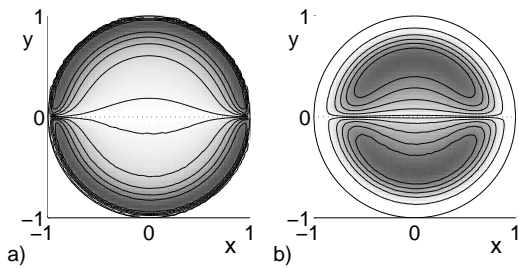}
\caption{\label{Fig2} Contour plot of the normalized density of complex eigenvalues $\rho_C(z)/(1-N_R)$ of truncations
of random orthogonal matrices for  a) $N=12$ and $M=10$ (weak non-orthogonality)
and b) $N=12$ and $M=6$  (strong nonorthogonality) with the larger values coded darker.
The density plotted in the upper half of the complex planes represents analytical results, while the density in the lower part is obtained numerically for a sample of $0.5\times 10^7$ random orthogonal matrices. }
\end{figure}

Now we shall look at complex eigenvalues. Their density $\rho_C(z)$, visualized in Fig.~\ref{Fig2}, vanishes on the real line. In the limit of strong non-orthogonality, close to the real line the complex density is described by the same scaling law
\[
\rho_C(z)\! \simeq\! \rho_R (x)^2 h(y\rho_R(x)), \ h(y)\!=\! 4\pi |y|
\rme^{4\pi y^2}\! \erfc(\sqrt{4\pi} |y|),
\]
as in the real Ginibre ensemble \cite{Edelman97}, while away from the real axis $\rho_C (z)$ is the same as for truncations of random \emph{unitary} matrices \cite{ZS00}, $ \rho_C (z)\simeq \frac{L}{\pi}\, \frac{1}{(1-|z|^2)^2}\, \Theta(\frac{M}{N} - |z|^2)$. In the limit of weak non-orthogonality away from the real axis
\[
\rho_C (z)\simeq \nu^2 h(4\pi \nu (1-|z|)), \ h(u)\!=\!4\pi L u^{L-1} \gamma^*(L+1, u),
\]
where $\nu=\frac{M}{2\pi}$ is the density of distribution of eigenvalues of random orthogonal matrices along the unit circle. Again, the limiting density profile does not depend on the angle and is the same as for truncations of random unitary matrices \cite{ZS00}.

Eigenvalue correlations for truncations of random orthogonal matrices can also be obtained in a closed form. In the limit of strong non-orthogonality, after appropriate rescaling the eigenvalue correlations become identical to those in the real Ginibre ensemble. For example, at the origin,  $f^2({z}/{\sqrt{L}})\simeq \sqrt{{L}/{(8\pi)}}\rme^{y^2-x^2}\erfc (\sqrt{2}|y|)$ and $K({z_1}/{\sqrt{L}}, {z_2}/{\sqrt{L}})\simeq \frac{1}{\sqrt{L}}(z_1-z_2)\rme^{z_1z_2}$ which is what one gets in the real Ginibre ensemble \cite{Som07,FN07,BS09}. The limit of weak non-orthogonality away from the real axis the eigenvalue correlations for truncations of random orthogonal matrices are exactly the same as those found for truncated random unitary matrices \cite{ZS00,KS09}. New correlation laws arise in the vicinity of $x=\pm1$.

\squeezetable
\begin{table}[h]
\caption{Ensembles of random nonhermitian matrices}
  \smallskip
{\renewcommand{\arraystretch}{2.67}
\begin{tabular}
[c]{c c c  c}\hline \hline
matrices & complex & real &  $\mu=M/N$ \\
\hline
 \parbox{2.8cm}{\centering Haar \\ measure } & $U(N) $ & $O(N)$  & $  \mu=1$ \\
\parbox{2.8cm}{\centering truncations of \\ matrices}
 & unitary  & orthogonal  & $1> \mu >0$\\
\parbox{2.8cm}{\centering Ginibre  \\ ensemble}
& complex & real &  $\mu \to 0$ \\
\hline \hline
\end{tabular}
}
\label{tab1}
\end{table}  

In conclusion, we have found the full probability distribution of eigenvalues of truncated random orthogonal matrices and obtained the eigenvalue density and higher order correlation functions in a closed form. 
This work completes our understanding of ensembles of non-hermitian random matrices which are derived from random unitary and orthogonal matrices. As shown in Table I truncations of  random orthogonal (unitary) matrices form an ensemble which interpolates between matrices distributed according to the Haar measure on the orthogonal (unitary) group \cite{Me91} and the real (complex) Ginibre ensemble \cite{Gi65}.

Financial support by the SFB Transregio-12 project
der Deutschen Forschungsgemeinschaft and
the special grant number DFG-SFB/38/2007 of
Polish Ministry of Science and Higher Education
is gratefully acknowledged.

Note added. After submitting our work we learned about a very recent preprint
of Forrester \cite{F10} extending our study


\begin{thebibliography}{28}
\bibitem{ftn} This distribution depends on the size $M$ of the truncation but not on 
its position in view of the invariance of the Haar measure.
\end{thebibliography}

\end{document}